# Determination of Al occupancy and local structure for β-$(Al_xGa_{1-x})_2O_3$ alloys across nearly full composition range from Rietveld analysis

Jayanta Bhattacharjee[1, 2], Archna Sagdeo[1, 2], and S. D. Singh[1, 2, a)]

[1]Synchrotrons Utilization Section, Raja Ramanna Centre for Advanced Technology, Indore-452013, India

[2]Homi Bhabha National Institute, Anushkati Nagar, Mumbai-400094, India

## Abstract

Al occupancy and local structure (bond length and bond angles) for monoclinic β-$(Al_xGa_{1-x})_2O_3$ alloys, with Al compositions (x) up to 90%, have been determined from Rietveld refinement of x-ray diffraction data. Al atom preferentially occupies octahedron ($O_h$) atomic site in comparison to tetrahedron ($T_d$) atomic site. However, sizable number of $T_d$ atomic sites i.e. 20% for Al composition of 5% remain occupied by Al atoms, which is found to increase sharply with Al composition. The $O_h$ atomic sites are not fully occupied by Al atoms even for Al composition of 90%. The lattice parameters (band gap) of β-$(Al_xGa_{1-x})_2O_3$ alloy decrease (increases) linearly with Al composition, but a change in slope of variation of both lattice parameters and band gap is observed at around Al composition of 50%. The lattice is found to be distorted for Al compositions more than 50% as indicated by large change in the bond angles. The lattice distortion is determined to be the origin for the observed change in slope for the variation of both lattice parameters and band gap for monoclinic β-$(Al_xGa_{1-x})_2O_3$ alloy system. Our results provide an insight in to the local structure of β-$(Al_xGa_{1-x})_2O_3$ alloys, which are required to have better understanding of their physical properties.

a)Corresponding author, Email:devsh@rrcat.gov.in; shreyashkar@gmail.com (S. D. Singh)

## Introduction

Wider band gap (~4.8 eV), higher dielectric constant (~10), larger breakdown voltage (~13 MV/cm), moderate electron mobility (~300 $cm^2$/V-s), n-type dopability and availability of good quality single crystal make β-$Ga_2O_3$ quintessential candidate for power electronic, deep ultraviolet (DUV) opto-electronic device etc. applications. [1-14] Al substituted β-$Ga_2O_3$ *i.e.* β-$(Al_xGa_{1-x})_2O_3$ alloy has gained tremendous attention because its band gap can be tailored from 4.5 eV to 6.9 eV. [15-19] Generally, maximum amount of Al composition (x) of 70% to 80% has been achieved experimentally either in the form of bulk or thin film by maintaining monoclinic phase for β-$(Al_xGa_{1-x})_2O_3$ alloys. [15-17, 20, 21] This fact can also be understood from Density functional theory (DFT) based calculations for β-$(Al_xGa_{1-x})_2O_3$ alloys, which revealed that the formation energy for monoclinic β phase is smaller than that of corundum α phase for x up to 70%, and thereafter corundum α phase was found to have lower formation energy. [19] In addition to this, the DFT calculations also predicted that the Al atom preferentially replaces Ga atom occupying octahedral ($O_h$) atomic sites as compared to that of tetrahedral ($T_d$) atomic sites for Al composition (x) ≤ 50%. Beyond this, Ga atom occupying the $T_d$ site is replaced by Al atom. However, the energy difference is only about 0.13 eV between the minimum energy configurations of Al atom replacing Ga atom occupying $O_h$ and $T_d$ sites. [19] Therefore, till date, it is believed that the Al atom replaces Ga atom sitting at $O_h$ site only for Al composition of x ≤ 50% and beyond this Al composition, it is the Ga occupying $T_d$ atomic site is replaced by the Al atom. In addition to this, there is hardly any experimental study available that determines the occupancy of Al atoms at $O_h$ and $T_d$ atomic sites in β-$(Al_xGa_{1-x})_2O_3$ alloys. Rietveld refinement of x-ray diffraction (XRD) data has been found a powerful tool to determine the occupancy of a substituent atom on the available atomic sites in the unit cell as well as other structural parameters. [22-25] Hence, in the present article, we carry out Rietveld refinement of XRD data of β-$(Al_xGa_{1-x})_2O_3$ alloys, which were synthesized by using standard solid-state reaction. The occupancy of Al atom at a particular atomic site, defined as the ratio of number of Al atoms occupy $O_h$ ($T_d$) atomic sites to the total number of Al atoms, has been quantified as a function of Al composition (x), where sizable number of Al atoms have been found to replace Ga atoms occupying $T_d$ atomic sites in addition to that occupying $O_h$ atomic sites for Al composition (x) ≤ 50%. However, $O_h$ atomic sites are preferred to be occupied by the Al atoms in the studied composition range. Additionally, the $O_h$ atomic sites are not fully occupied by the Al atoms even for Al compositions in range of 50% ≤ x ≤ 90%; however, occupancy of Al atom at the $O_h$ atomic site comes closer to that of fully replaced value with increase in Al composition. Further, the variation of lattice parameters (a, b, c, β, V) shows a change in slope at around the Al composition of 50%. Furthermore, the variation of band gap for β-$(Al_xGa_{1-x})_2O_3$ alloys also displays a change in its slope at same value of Al composition. The change of slope of band gap variation with Al composition has been ascribed in the literature to the existence of two band gap bowing parameters, one for Al composition (x) below 50% and other for x ≥ 50%. The observation of change in slope in the variation of lattice parameters and the band gap has been attributed to the distortion of the monoclinic β-$(Al_xGa_{1-x})_2O_3$ lattice, which is caused by the large variation in the bond angles for Al compositions, x ≥ 50%.

## Sample preparation and experimental details

β-$(Al_xGa_{1-x})_2O_3$ polycrystalline samples with different Al compositions (x) ranging from of 0 to 95% were prepared using conventional solid-state reaction method. The process of synthesis of β-$(Al_xGa_{1-x})_2O_3$ polycrystalline samples is similar as reported earlier. [16] XRD experiments on polycrystalline samples were carried out by using Ni filtered Cu Kα radiation lab source on a D8 ADVANCE, BRUKER diffractometer to determine phase purity and crystalline structure. The XRD data were collected in 2θ range of 10° - 80° with a step of 0.01° with an average time of 1.5 sec per step and by rotating the sample at a speed of 20 rpm. Rietveld refinement of measured XRD profile was carried by using FullProf software, [25] which provided the structural information like lattice parameters, bond lengths, bond angles and Al occupancy at $O_h$ and $T_d$ atomic sites of the β monoclinic unit cell.

## Results and discussions

Figure 1 shows selected XRD profiles collected from β-$(Al_xGa_{1-x})_2O_3$ polycrystalline samples, which have been vertically shifted for clarity of presentation. The XRD data from β-$(Al_xGa_{1-x})_2O_3$ samples can be catalogued to monoclinic β-phase by the Joint Committee for Powder Diffraction Standards (JCPDS; card no. 00-041-1103). The monoclinic β-phase is maintained up to 90% of Al substitution in β-$Ga_2O_3$ host material to form monoclinic β-$(Al_xGa_{1-x})_2O_3$ alloys. However, for Al composition of 95%, the XRD profile contains some new XRD peaks, marked by arrows in the graph, as shown in the inset to Fig. 1, which are identified to be related to α-$Al_2O_3$. Thus, β-$(Al_xGa_{1-x})_2O_3$ alloy is determined to be in single β-phase for Al compositions x ≤ 90%, which is the highest Al compositional β-$(Al_xGa_{1-x})_2O_3$ alloy synthesized under equilibrium experimental conditions. [15] Additionally, insets to Fig. 1 also show the shifting of Bragg peaks towards higher angle indicating the compression of unit cell with Al substitution, which is due to smaller size of Al atom in comparison to the Ga atom. Along with this, the XRD peak intensity decreases systematically with increasing Al composition, which is related to the lower atomic scattering factor of Al atom. [15,16]

Firstly, we schematically show a unit cell of $\beta$-$Ga_2O_3$ in Fig. 2(a), which has been drawn by using freely available VESTA software based on the atomic positions as mentioned by S. Geller[26] and obtained using Rietveld refinement in the present study. There are two inequivalent positions of Ga atom in the unit cell of $\beta$-$Ga_2O_3$, where one Ga atom, designated as Ga1, is coordinated in $T_d$ geometry and the other one, designated as Ga2, is coordinated in $O_h$ geometry. Besides this, there are three inequivalent positions of O atom in the unit cell of $\beta$-$Ga_2O_3$, designated as O1, O2, and O3. One O1 atom, two O2 atoms, and one O3 atom are bonded with Ga1 atom in the unit cell of $\beta$-$Ga_2O_3$. Further, two O1 atoms, one O2 atom, and three O3 atoms are bonded with Ga2 atom. Thus, the Al atom finds two atomic sites *i.e.* $T_d$ and $O_h$ to replace Ga atom in the unit cell of $\beta$-$Ga_2O_3$. Hence, to determine Al occupancy and other structural parameters, Rietveld refinement of XRD data of single phase monoclinic $\beta$-$(Al_xGa_{1-x})_2O_3$ alloys for Al composition up to 90% has been carried out using FullProf software. [25] Figure 2(b) shows the Rietveld fitting of XRD data for one of the $\beta$-$(Al_xGa_{1-x})_2O_3$ alloy sample for 10% of Al composition as a representative, where experimental data has been displayed by symbols. The lines in Fig. 2 display Rietveld fit to the experimental data by putting 100% Al atoms at $T_d$ atomic sites (dashed line), 0% Al atoms at $T_d$ atomic sites (dotted line), and 50% Al atoms at $T_d$ atomic sites (dashed-dotted line). It is clear from the Rietveld fitting, as shown in the inset to Fig. 2, where enlarged views at some Bragg peaks are shown, that whole of the Al atom neither fully replaces Ga atom occupying $T_d$ atomic site nor occupying $O_h$ atomic site. The fitting is somewhat reasonable if 50% $T_d$ and 50% $O_h$ atomic sites are occupied by the Al atom. Lastly, the best fit is noted if 26% $T_d$ and 74% $O_h$ atomic sites occupied by the Al atom in the unit cell of $\beta$-$(Al_xGa_{1-x})_2O_3$ alloy. Apart from this, a mismatch in intensity between the fitted and the experimental data for (002) Bragg peak is observed and this observation is noted for all samples except pristine $\beta$-$Ga_2O_3$. This fact may be due to the preferred orientation of grains along [001] direction for $\beta$-$(Al_xGa_{1-x})_2O_3$ alloys. Please see Supplementary information for details on the Rietveld fitting of all other samples. Parameters indicating the quality of Rietveld fit for all the samples have been listed in the Table S1 (as shown in Supplementary information), where it is noted that the Rietveld fitting is quite good for Al compositions up to 70%. However, the Rietveld fitting of samples for Al compositions of 80% and 90% are relatively poorer as shown in Figs. S10 and S11 (Supplementary information), which will be discussed subsequently.

The value of Al occupancy (symbols) at $O_h$ and $T_d$ atomic sites as obtained from the Rietveld fitting for all β-$(Al_xGa_{1-x})_2O_3$ alloys has been displayed in Fig. 3(a). The dotted-dashed line indicates the value of Al occupancy, when Al atom replaces $T_d$ and $O_h$ sites in equal proposition *i.e.* 50%. The dashed lines give the value of Al occupancy, when Al atom replaces only Ga atom sitting at $O_h$ site for Al compositions, x ≤ 50% and thereafter, only Ga atom occupying $T_d$ atomic site is replaced by Al atom. This was proposed by the DFT calculations as reported in the literature. [18, 19] However, it is interesting to note that sizable amount of Al atoms does replace Ga atoms occupying $T_d$ sites in addition to that occupying $O_h$ sites. The Al occupancy at $T_d$ ($O_h$) atomic sites sharply increases (decreases) with increase in Al compositions up to 30% and then it does not change much with further increase in Al composition from 30% to 50%. Thereafter, Al occupancy at $T_d$ ($O_h$) atomic site again increases (decreases) with increase in Al composition from 50% to 90%. Hence, for β-$(Al_xGa_{1-x})_2O_3$ alloys, it is very clear that the Al atom does occupy significant number of $T_d$ sites even for Al compositions, x ≤ 50% and tendency of Al atoms occupying the $T_d$ sites in comparison to the $O_h$ sites increases with the Al composition. This is new and important addition to the existing knowledge, where it was believed that only $O_h$ atomic sites are filled for Al composition x ≤ 50% and then, $T_d$ sites are filled for Al compositions larger than 50%. [15, 18, 19] Occupancy of sizable number of $T_d$ atomic sites by Al atoms for lower Al compositional β-$(Al_xGa_{1-x})_2O_3$ alloys is not surprising, because there is not a large difference (0.13 eV) between the minimum energy configurations of Al occupying $O_h$ and $T_d$ atomic sites as predicted by DFT calculations.[19] Further, it is also important to note that the $O_h$ sites are not fully occupied by the Al atoms even for Al composition of 90%.

Next, we investigate the variation of lattice parameters (a, b, c, β, V) with Al compositions as displayed in Figs. 3(b)-3(f). We observe that the values of lattice parameters a, b, c, and V decrease linearly with Al composition, while, the value of monoclinic tilt angle *β* is found to increase linearly with the Al composition. However, a change in slope is observed in the variation of these lattice parameters at around Al composition of 50%. The value of slope of variation of the lattice parameters a, b, c, β, and V are determined to be -0.0032 ± 0.0001 (-0.0011 ± 0.0001), -0.00099 ± 0.00002 (-0.00033 ± 0.00003), -0.0013 ± 0.0005 (-0.0004 ± 0.00006), 0.0030 ± 0.00006 (0.0007 ± 0.0001), and -0.17 ± 0.004 (-0.056 ± 0.006), respectively

for Al composition $x \leq 0.50$ ($x \geq 0.50$). The origin of change in the slope values at around Al composition of 50% in the variation of lattice parameters will be discussed subsequently.

The variation of bond lengths like Ga1-O1, Ga1-O2, Ga1-O3 forming the tetrahedron with Ga1 and Ga2-O1, Ga2-O2, Ga2-O3, and Ga2-O3$^+$ forming the octahedron with Ga2 within the β-$(Al_xGa_{1-x})_2O_3$ unit cell are displayed in the left column of Fig. 4. The Ga2-O3$^+$ is the bond length, which joins two Ga2 atoms along b direction and this is larger than the other Ga2-O3 bond length. Similarly, the variation of bond angles belonging to octahedron-oxygen-octahedron ($O_h$-O-$O_h$), tetrahedron-oxygen-tetrahedron ($T_d$-O-$T_d$) and tetrahedron-oxygen-octahedron ($T_d$-O-$O_h$) system are shown in the right column of Fig. 4. The bond angles Ga1-O3-Ga2$^*$ and Ga2-O3-Ga2$^*$ are those bond angles, where both Ga atoms are in the same plane. The general observation is that both bond lengths and bond angles slowly vary with Al compositions up to 50% and thereafter, they both display a rapid change with further increase in the Al composition. Thus, it is inferred that the lattice parameters (a, b, c) and volume of β-$(Al_xGa_{1-x})_2O_3$ unit cell decrease with Al compositions up to 50%, while keeping a minimal lattice distortion as the change in the bond angles is quite small (2-3° maximum). On the other, for larger (≥ 50%) Al compositions, lattice parameters (a, b, c) and volume (V) of the unit cell do decrease with Al composition, but large change (~12° maximum) in the bond angles make the β-$(Al_xGa_{1-x})_2O_3$ lattice to be highly distorted. The large distortion in the lattice causes the lattice parameter and volume to vary at slow rate for Al compositions larger than 50%. Thus, lattice distortion caused by the large change in the bond angles is responsible for the change in the slope of the lattice parameter variation at around Al composition of 50%. It is also interesting to note that larger change in the bond lengths occurs for samples with Al compositions of 80% and 90% indicating maximum lattice distortion in these samples, which is the reason that the parameters indicating the quality of Rietveld fit for these samples are relatively poorer as mentioned earlier. Additionally, it is the highly distorted nature of the monoclinic unit cell of β-$(Al_xGa_{1-x})_2O_3$ alloy, indicated by large change in the bond angles, prevents further incorporation of Al atom beyond 90% in equilibrium experimental conditions.

Another consequence of the lattice distortion of a material is the change in its band gap, which is mainly governed by the overlap of atomic orbitals. [27] The overlap of atomic orbitals gets affected due to large change in the bond angles leading to change in the band gap value. Figure 5 displays the variation of band gap of β-$(Al_xGa_{1-x})_2O_3$ alloys, as determined from 10 K optical reflectivity experiments mentioned elsewhere, [16] with Al composition. It is quite interesting to note that a change in the slope of the variation of the band gap of the alloys is also noted at around Al composition of 50%. It is noteworthy to recall that a change in the slope of variation of lattice parameters is noted as presented in the Figs. 3(b)-3(f). Thus, change in slope of the band gap variation of β-$(Al_xGa_{1-x})_2O_3$ alloys confirms that the lattice becomes highly distorted for Al compositions greater than 50%. Thus, the lattice distortion for high Al compositional β-$(Al_xGa_{1-x})_2O_3$ alloys is found to be responsible for the existence of two bowing parameters for these alloys as reported in the literature.[17, 18]

## Conclusion

In conclusion, we have studied the local structure of β-$(Al_xGa_{1-x})_2O_3$ alloys from Rietveld refinement of XRD data. β-$(Al_xGa_{1-x})_2O_3$ alloys are found to be of single phase for Al compositions up to 90%. Sizable amount of Al atom is found to populate $T_d$ atomic sites in addition to $O_h$ atomic sites for Al compositions lesser than 50%. Occupancy of Al atom at $T_d$ atomic sites relative to $O_h$ sites is found to sharply increase with Al compositions up to 30%, then it does not change much for Al compositions up to 50%. Thereafter, it increases slowly with further increase in Al composition. The $O_h$ sites are still not fully occupied with the Al atoms even for Al composition of 90%. Although, the $O_h$ atomic sites are preferred to be occupied by the Al atoms in the whole composition range of 90%. The lattice parameters are found to vary linearly with a change in slope at around Al composition of 50%, which has been attributed to the lattice distortion caused by the large variation in the bond angles. The lattice distortion is also found to be the reason of the change in the slope of the band gap variation.

## Acknowledgement

The authors acknowledge Dr. Tapas Ganguli, Head, SUS, Dr. Ravindra Jangir, Dr. Satish Mandal, Shri Ashok Kumar, Dr. Sahadeb Ghosh, Dr. Preeti Pokhriyal and Shri Shankar Dutt for fruitful discussions and Dr. S. V. Nakhe, Director, RRCAT, Indore for his constant support during the course of this work. Jayanta Bhattacharjee acknowledges his Doctoral committee members for the useful suggestions and HBNI, RRCAT for the support.

**References**

[1]R. Roy, V. G. Hill, and E. F. Osborn, **J. Am. Chem. Soc. 74**,719 (1952).

[2]S. J. Pearton, J. Yang, P. H. Cary, F. Ren, J. Kim, M. J. Tadjer, and M. A. Mastro, **Appl. Phys. Rev. 5**, 011301 (2018).

[3]G. Yang, S. Jang, F. Ren, S. J. Pearton, and J. Kim, **ACS Appl. Mater. Interfaces 9**, 40471 (2017).

[4]M. H. Wong, A. Takeyama, T. Makino, T. Ohshima, K. Sasaki, A. Kuramata, S. Yamakoshi, and M. Higashiwaki, **Appl. Phys. Lett. 112**, 023503 (2018).

[5]Z. Liu, J. Yu, P. Li, X. Wang, Y. Zhi, X. Chu, X. Wang, H. Li, Z. Wu, and W. Tang, **J. Phys. D: Appl. Phys. 52**, 295104 (2019).

[6]J. Kim, S. J. Pearton, C. Fares, J. Yang, F. Ren, S. Kim, and A. Y. Polyakov, **J. Mater. Chem. C 7**, 10 (2019).

[7]Y. Kokubun, K. Miura, F. Endo, and S. Nakagomi, **Appl. Phys. Lett. 90,** 031912 (2007).

[8]M. Orita, H. Ohta, M. Hirano, and H. Hosono, **Appl. Phys. Lett. 77,** 4166 (2000).

[9]M. Higashiwaki, K. Sasaki, A. Kuramata, T. Masui, and S. Yamakoshi, **Appl. Phys. Lett. 100,** 013504 (2012).

[10]E. Chikoidze, T. Tchelidze, C. Sartel, Z. Chi, R. Kabouche, I. Madaci, C. Rubio, H. Mohameda, V. Sallet, F. Medjdoub, A. Perez-Tomas, Y. Dumont, **Materials Today Physics 15**, 100263 (2020).

[11]H. H. Tippins, **Phys. Rev. 140,** A316 (1965).

[12]T. Motsumoto, M. Aoki, A. Kinoshita, and T. Aono, **Jpn. J. Appl. Phys. 13,** 1578 (1974).

[13]N. Ueda, H. Hosono, R. Waseda, and H. Kawazoe, **Appl. Phys. Lett. 71,** 933 (1997).

[14]J. Zhang, J. Shi, D. C. Qi, L. Chen, and K. H. L. Zhang, **APL Mater. 8,** 020906 (2020).

[15]B. W. Krueger, C. S. Dandeneau, E. M. Nelson, S. T. Dunham, F. S. Ohuchi, and M. A. Olmstead, **J. Am. Ceram. Soc. 99,** 2467 (2016).

[16]J. Bhattacharjee, S. Ghosh, P. Pokhriyal, R. Gangwar, R. Dutt, A. Sagdeo, P. Tiwari, S. D. Singh, **AIP Advances 11,** 075025 (2021).

[17]R. Wakabayashi, K. Yoshimatsu, M. Hattori, J. S. Lee, O. Sakata, and A. Ohtomo, **Cryst. Growth Des. 21,** 2844 (2021).

[18]J. B. Varley, **Journal of Materials Research. 36**, 4790 (2021).

[19]H. Peelaers, J. B. Varley, J. S. Speck, and C. G. Van de Walle, **Appl. Phys. Lett. 112,** 242101 (2018).

[20]C.-H. Liao, K.-H. Li, C. G. Torres-Castanedo, G. Zhang, and X. Li, **Appl. Phys. Lett. 118,** 032103 (2021).

[21]J. Li, X. Chen, T. Ma, X. Cui, F.-F. Ren, S. Gu, R. Zhang, Y. Zheng, S. P. Ringer, L. Fu, H. H. Tan, C. Jagadish, and J. Ye, **Appl. Phys. Lett. 113,** 041901(2018).

[22]H. M. Rietveld, **Acta Cryst. 22,** 151 (1967).

[23]R. J. Hill, **J. Appl. Cryst. 25**, 589 (1992).

[24]D. Mondal, C. Kamal, S. Banik, A. Bhakar, A. Kak, G. Das, V. R. Reddy, A. Chakrabarti, and T. Ganguli, **J. Appl. Phys. 120**, 165102 (2016).

[25]J. Rodriguez-Carvajal, **Physica B 192**, 55 (1993).

[26]S. Geller, J. Chem. Phys. 33, 676 (1960).

[27]J. E. Bernard and A. Zunger, **Phys. Rev. B 34**, 5992 (1986).

## Figure Captions

**Figure 1** X-ray diffraction (XRD) profiles of selected β-$(Al_xGa_{1-x})_2O_3$ alloys (x=0%, 10%, 35%, 50%, 80%, 90%, and 95%) taken with Cu Kα radiation. The XRD profiles have been shifted vertically for clarity of presentation. The XRD pattern for pristine β-$(Ga)_2O_3$ has been indexed with JCPDS card no. 00-041-1103. Inset to Fig. displayed an enhanced view around 2θ values of 15° and 38°, where Bragg peaks are seen to be shifted towards larger angles with increase in the Al composition. Left inset shows (001) and (-201) Bragg reflections, which are found to shift towards larger angles with increases in Al composition. A secondary phase of α-$Al_2O_3$ has been observed as marked by arrows in the right inset.

**Figure 2 (a)** Schematic of unit cell of β-$Ga_2O_3$ indicating atomic positions of Ga and O atoms drawn from using freely available VESTA software. **(b)** x-ray diffraction (XRD) profile of β-$(Al_{0.1}Ga_{0.9})_2O_3$ sample, where symbols indicate experimental data. Solid lines are the Rietveld fitted data for various occupancy of Al atoms at tetrahedral ($T_d$) and octahedral ($O_h$) sites. Insets present the enhanced view at certain Bragg peaks showing the quality of fitting.

**Figure 3 (a)** Occupancy of Al atom at tetrahedral ($T_d$) and octahedral ($O_h$) atomic sites, determined from the Rietveld fitting, as a function of Al composition of β-$(Al_xGa_{1-x})_2O_3$ alloys. Dashed lines indicate the value of Al occupancy, when only $O_h$ atomic sites are filled by Al atoms for Al composition x ≤ 50% and thereafter, only $T_d$ atomic sites are filled by Al atoms as reported by Density functional theory in literature. Dashed-dotted line indicates the value of Al occupancy, when Al atom fills $O_h$ and $T_d$ atomic sites in equal proportion. Variation of lattice parameters **(b)** a, **(c)** b, **(d)** c, **(e)** β, and **(f)** volume (V) as a function of Al composition. Symbols display values of lattice determined XRD data. Dotted lines are the linear fitting indicating two slopes.

**Figure 4** Bond lengths forming tetrahedral and octahedral bonding arrangement of unit cell of β-$(Al_xGa_{1-x})_2O_3$. Bond angles between tetrahedron-oxygen-tetrahedron ($T_d$-O-$T_d$), octahedron-oxygen-octahedron ($O_h$-O-$O_h$), and tetrahedron-oxygen-octahedron ($T_d$-O-$O_h$). Both bond length and bond angles are determined from Rietveld fitting of XRD data for different Al composition.

**Figure 5** Band gap of β-$(Al_xGa_{1-x})_2O_3$ alloys determined at 10 K from optical reflectivity data. Symbols represent the experimentally determined band gap, while dotted lines are the linear fitting. Values of slope of the linear fitting have also been provided. Arrow indicates a change in the slope at around Al composition of 50%.

**Fig. 1**

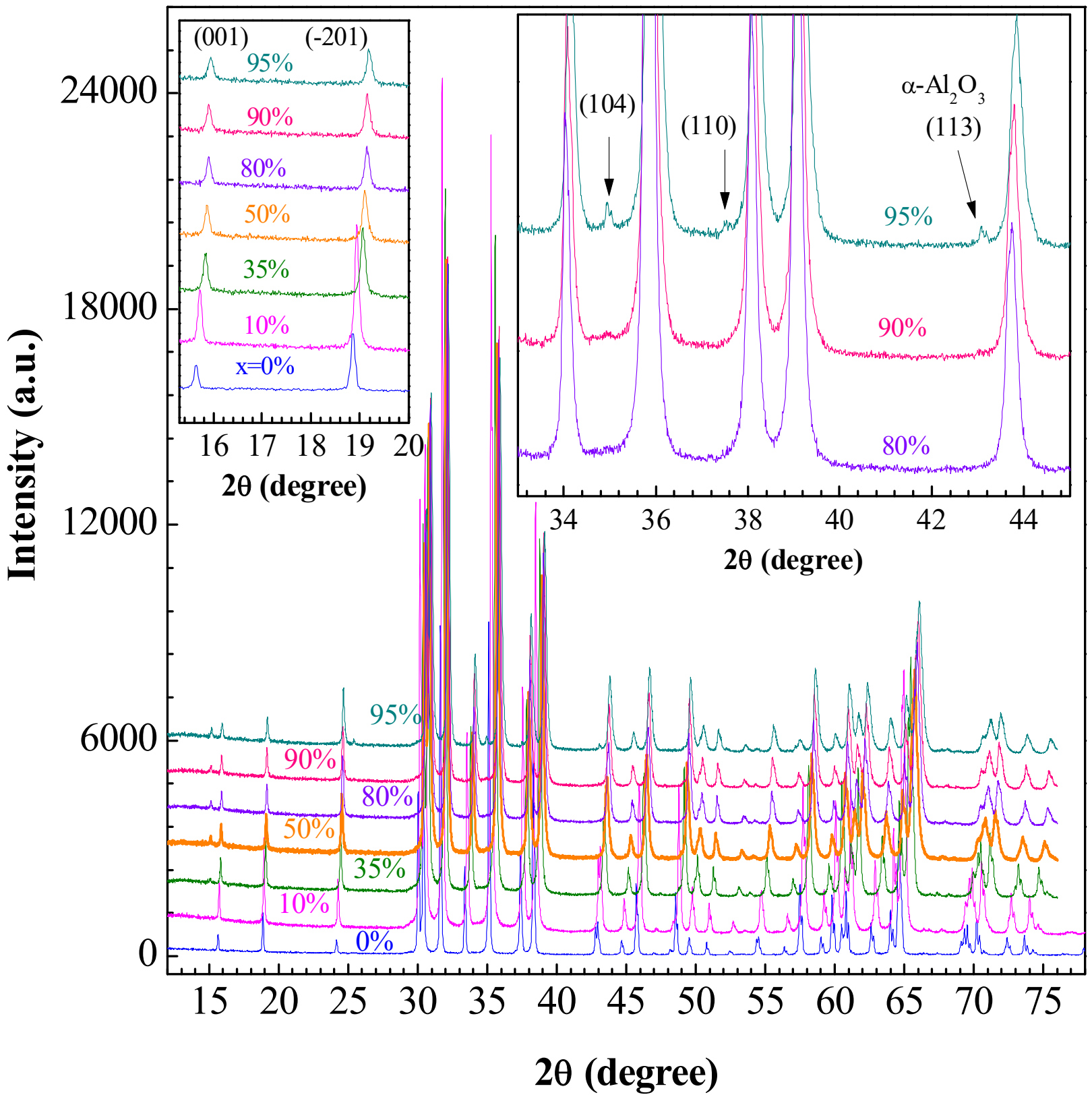

**Figure 1** X-ray diffraction (XRD) profiles of selected β-$(Al_xGa_{1-x})_2O_3$ alloys (x=0%, 10%, 35%, 50%, 80%, 90%, and 95%) taken with Cu Kα radiation. The XRD profiles have been shifted vertically for clarity of presentation. The XRD pattern for pristine β-$(Ga)_2O_3$ has been indexed with JCPDS card no. 00-041-1103. Inset to Fig. displayed an enhanced view around 2θ values of 15° and 38°, where Bragg peaks are seen to be shifted towards larger angles with increase in the Al composition. Left inset shows (001) and (-201) Bragg reflections, which are found to shift towards larger angles with increases in Al composition. A secondary phase of α-$Al_2O_3$ has been observed as marked by arrows in the right inset.

**Fig. 2**

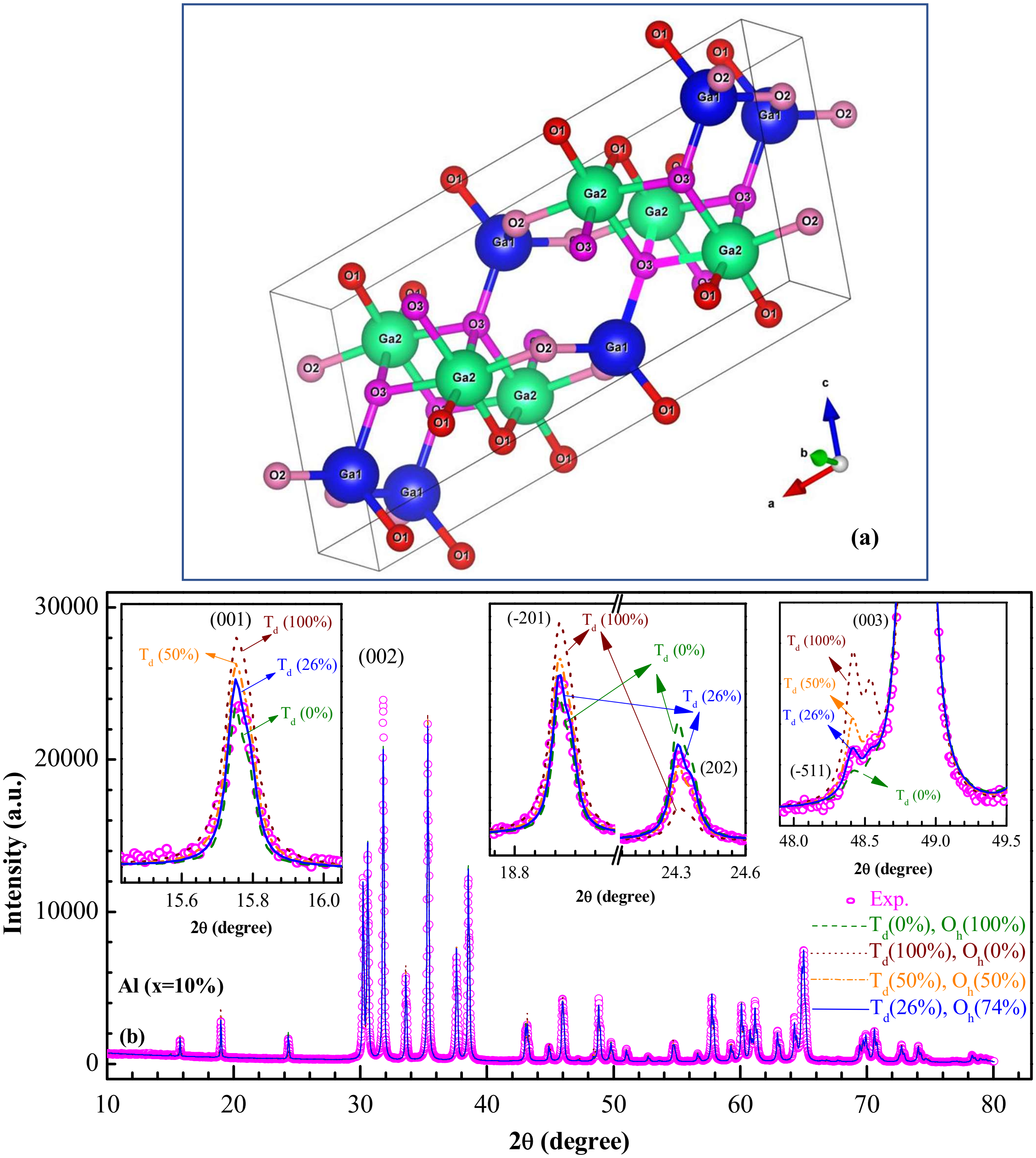

**Figure 2 (a)** Schematic of unit cell of β-$Ga_2O_3$ indicating atomic positions of Ga and O atoms drawn from using freely available VESTA software. **(b)** x-ray diffraction (XRD) profile of β-$(Al_{0.1}Ga_{0.9})_2O_3$ sample, where symbols indicate experimental data. Solid lines are the Rietveld fitted data for various occupancy of Al atoms at tetrahedral ($T_d$) and octahedral ($O_h$) sites. Insets present the enhanced view at certain Bragg peaks showing the quality of fitting.

**Fig. 3**

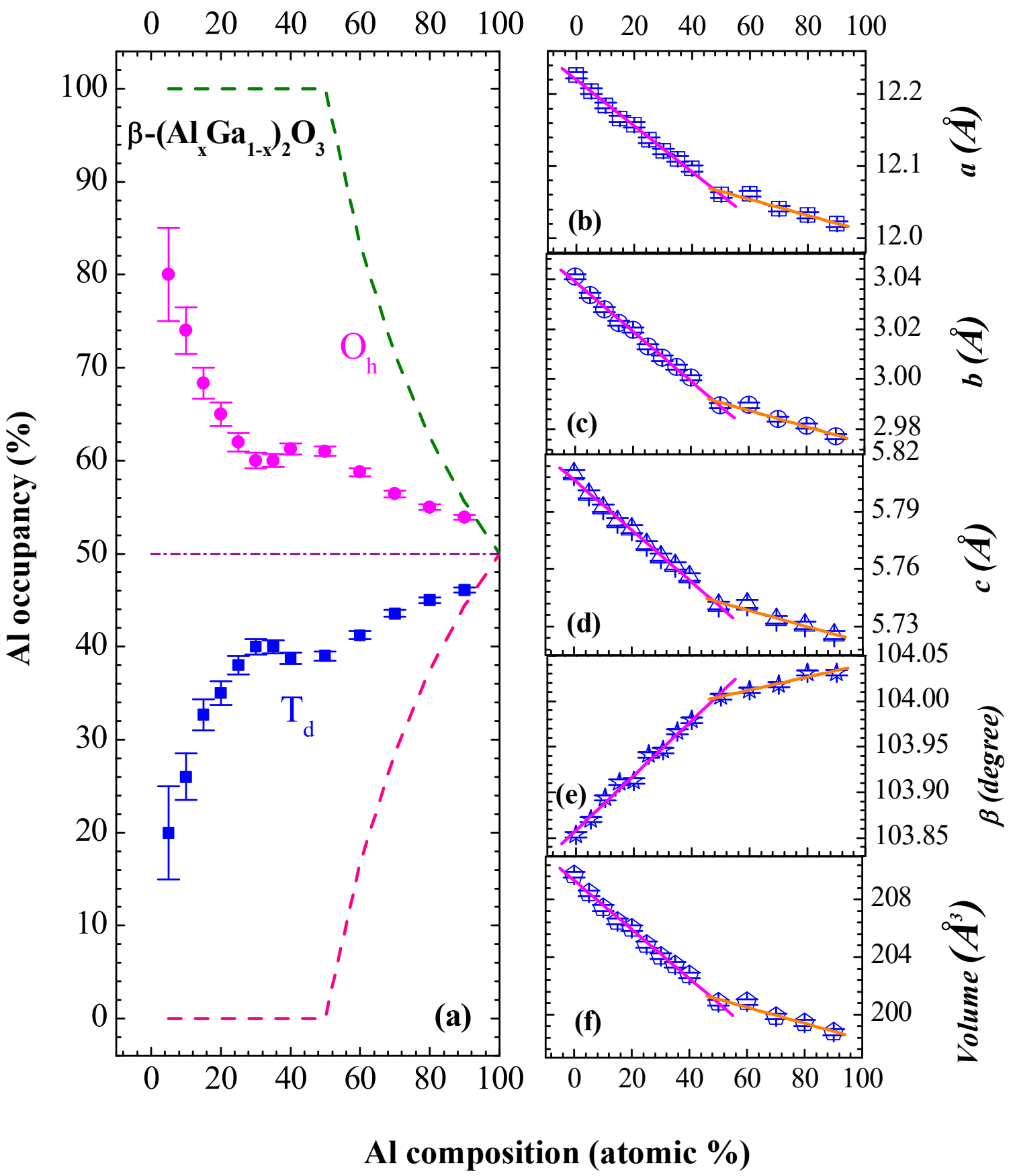

**Figure 3 (a)** Occupancy of Al atom at tetrahedral ($T_d$) and octahedral ($O_h$) atomic sites, determined from the Rietveld fitting, as a function of Al composition of β-$(Al_xGa_{1-x})_2O_3$ alloys. Dashed lines indicate the value of Al occupancy, when only $O_h$ atomic sites are filled by Al atoms for Al composition x ≤ 50% and thereafter, only $T_d$ atomic sites are filled by Al atoms as reported by Density functional theory in literature. Dashed-dotted line indicates the value of Al occupancy, when Al atom fills $O_h$ and $T_d$ atomic sites in equal proportion. Variation of lattice parameters **(b)** a, **(c)** b, **(d)** c, **(e)** β, and **(f)** volume (V) as a function of Al composition. Symbols display values of lattice determined XRD data. Dotted lines are the linear fitting indicating two slopes.

**Fig. 4**

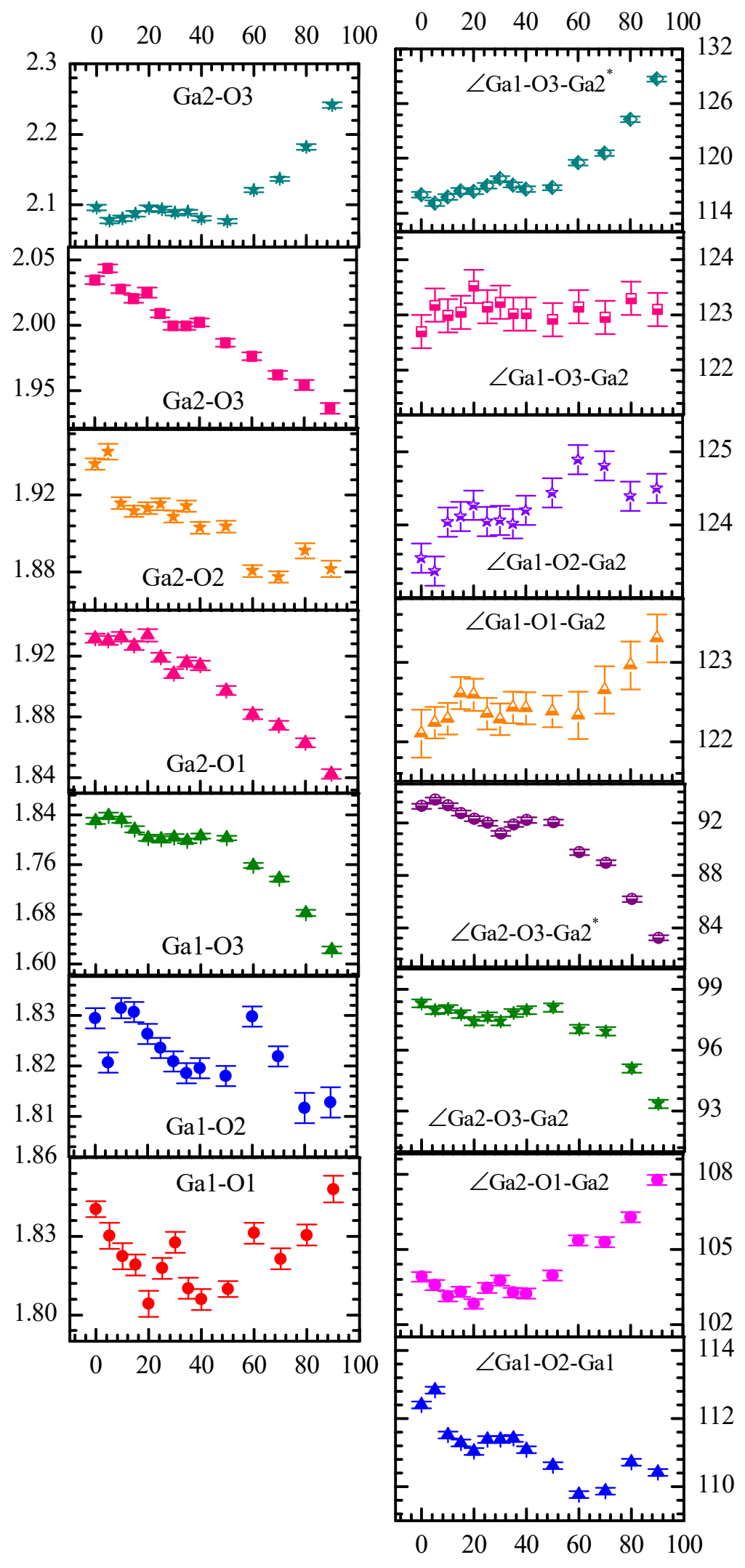

**Figure 4** Bond lengths forming tetrahedral and octahedral bonding arrangement of unit cell of β-$(Al_xGa_{1-x})_2O_3$. Bond angles between tetrahedron-oxygen-tetrahedron ($T_d$-O-$T_d$), octahedron-oxygen-octahedron ($O_h$-O-$O_h$), and tetrahedron-oxygen-octahedron ($T_d$-O-$O_h$). Both bond length and bond angles are determined from Rietveld fitting of XRD data for different Al composition.

**Fig. 5**

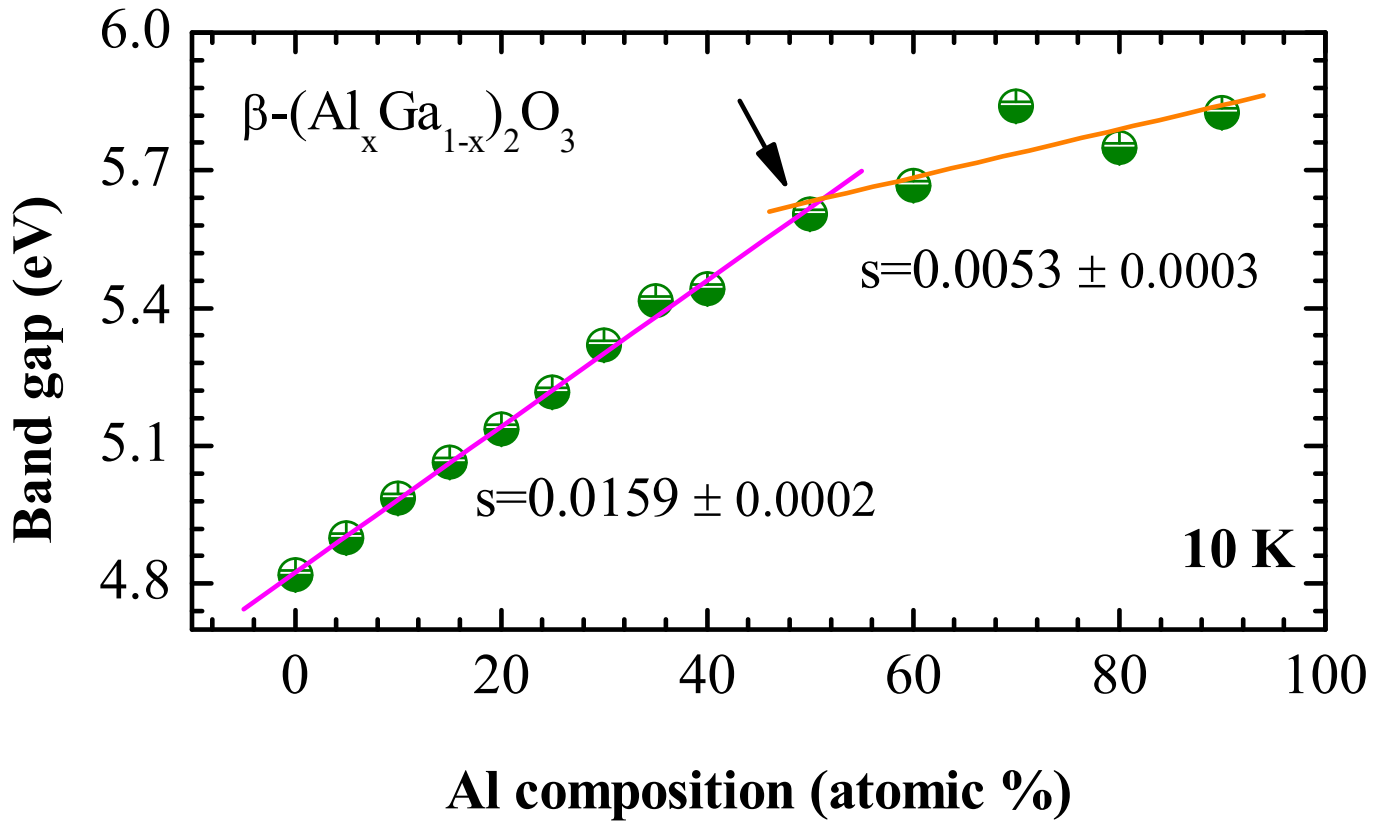

**Figure 5** Band gap of β-($Al_xGa_{1-x}$)$_2O_3$ alloys determined at 10 K from optical reflectivity data. Symbols represent the experimentally determined band gap, while dotted lines are the linear fitting. Values of slope of the linear fitting have also been provided. Arrow indicates a change in the slope at around Al composition of 50%.

## Supplementary information

**Table S1**

| S. N. | Al composition (x, %) | $R_p$ | $R_{wp}$ | $R_{exp}$ | $\chi^2$ |
|---|---|---|---|---|---|
| 1. | 0 | 10.2 | 12.8 | 7.9 | 2.61 |
| 2. | 5 | 8.17 | 9.58 | 4.58 | 4.37 |
| 3. | 10 | 8.37 | 9.75 | 4.74 | 4.23 |
| 4. | 15 | 8.17 | 9.57 | 4.73 | 4.10 |
| 5. | 20 | 5.69 | 9.98 | 4.87 | 4.19 |
| 6. | 25 | 8.55 | 9.84 | 4.76 | 4.26 |
| 7. | 30 | 8.63 | 9.73 | 4.97 | 3.83 |
| 8. | 35 | 8.52 | 9.80 | 4.85 | 4.08 |
| 9. | 40 | 8.98 | 9.86 | 5.03 | 3.84 |
| 10. | 50 | 7.45 | 8.70 | 4.79 | 3.31 |
| 11. | 60 | 9.38 | 11.0 | 5.09 | 4.64 |
| 12. | 70 | 9.86 | 11.9 | 5.16 | 5.33 |
| 13. | 80 | 12.2 | 14.8 | 5.22 | 8.09 |
| 14. | 90 | 14 | 18.3 | 4.67 | 15.3 |

**Table S1** Fit parameters indicating quality of Rietveld fitting for all samples.

**Fig. S1**

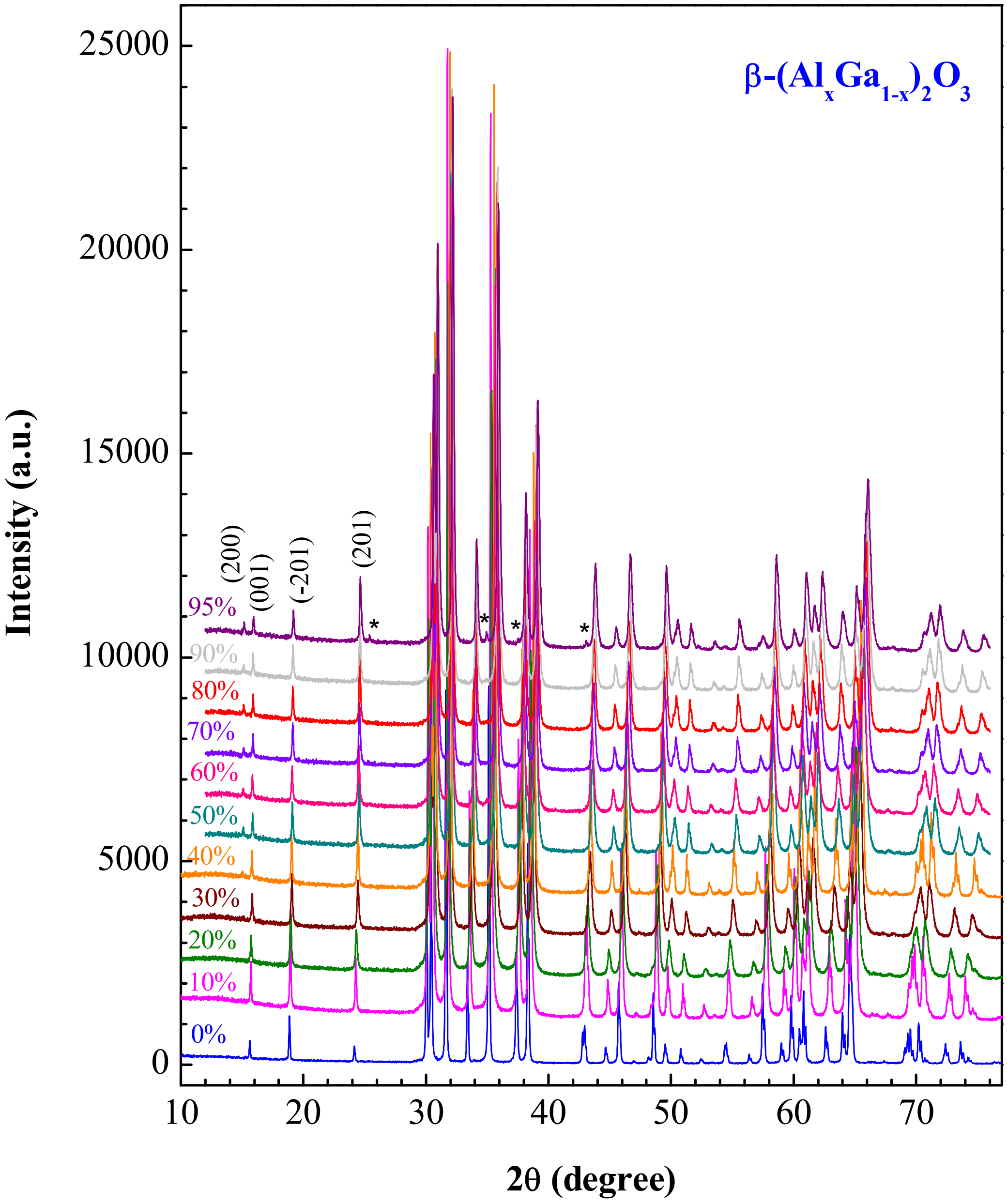

**Figure S1** X-ray diffraction (XRD) profiles for different β-$(Al_xGa_{1-x})_2O_3$ samples with Al compositions (x) varying from 0% to 95%. The XRD profiles have been shifted along vertical direction for clarity of presentation. Some of the Bragg reflections of β-$(Al_xGa_{1-x})_2O_3$ have been mentioned. The symbols * shows XRD peaks from the secondary phase of α-$Al_2O_3$, which are observed in XRD profile of β-$(Al_xGa_{1-x})_2O_3$ with Al composition of 95%.

**Fig. S2**

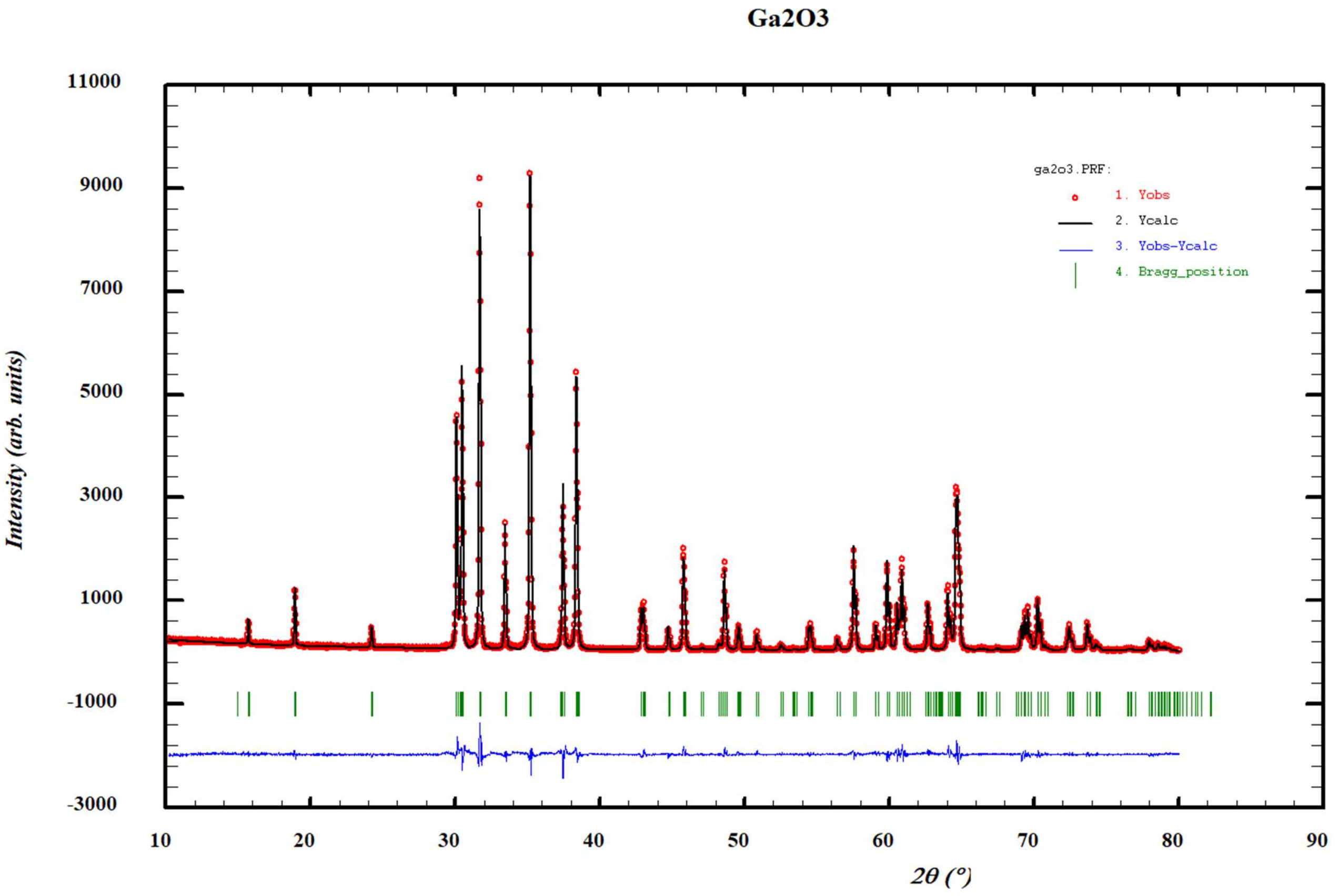

**Figure S2** Rietveld fitting of powder XRD data of β-$Ga_2O_3$. File generated from WinPlotr. Symbols indicate experimental XRD data. Red line is the Rietveld fitted data. Green lines indicate the positions of Bragg peaks. Blue line is the difference between experimental and fitted data.

**Fig. S3**

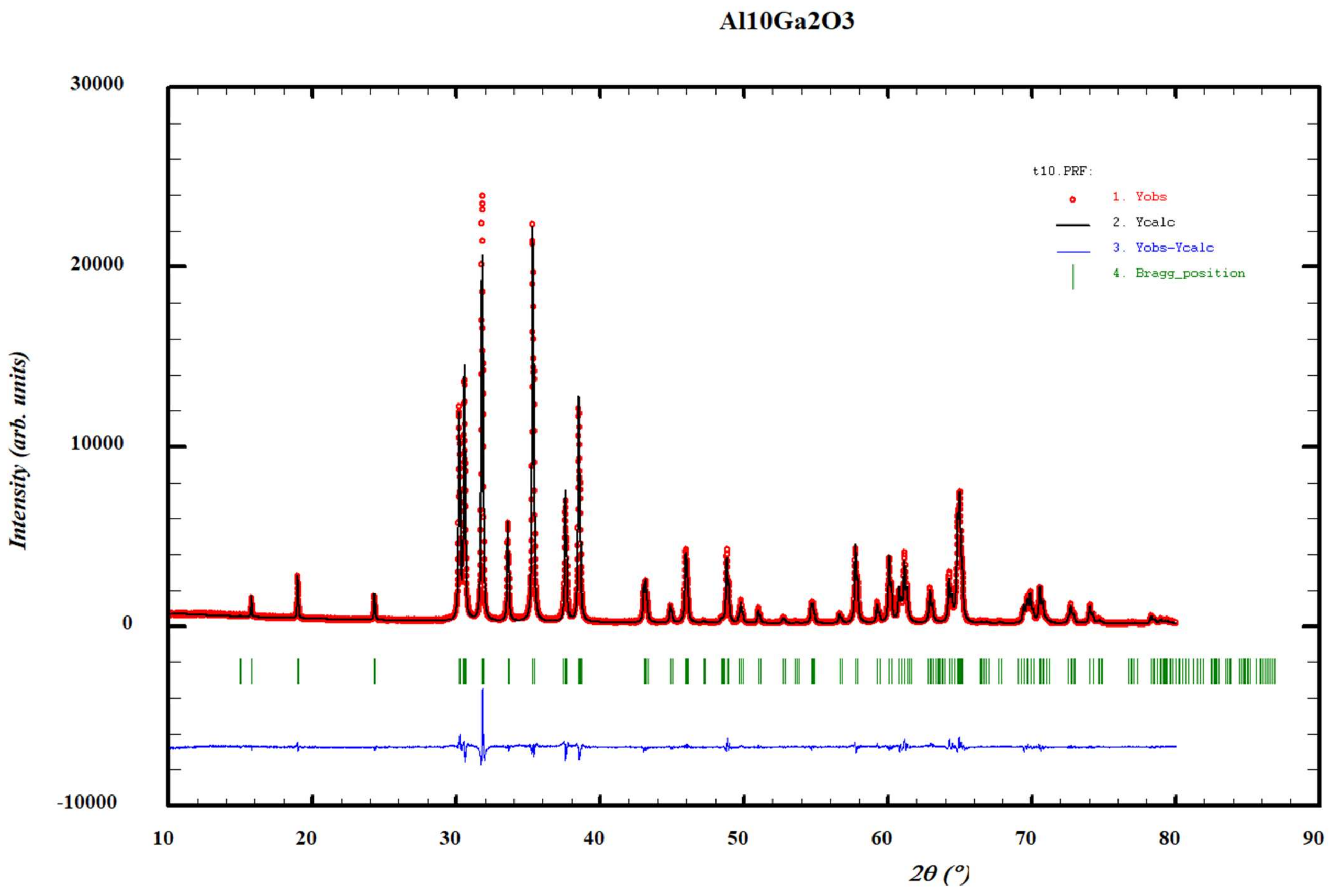

**Figure S3** Rietveld fitting of powder XRD data of β-$(Al_{0.1}Ga_{0.9})_2O_3$. File generated from WinPlotr. Symbols indicate experimental XRD data. Red line is the Rietveld fitted data. Green lines indicate the positions of Bragg peaks. Blue line is the difference between experimental and fitted data.

**Fig. S4**

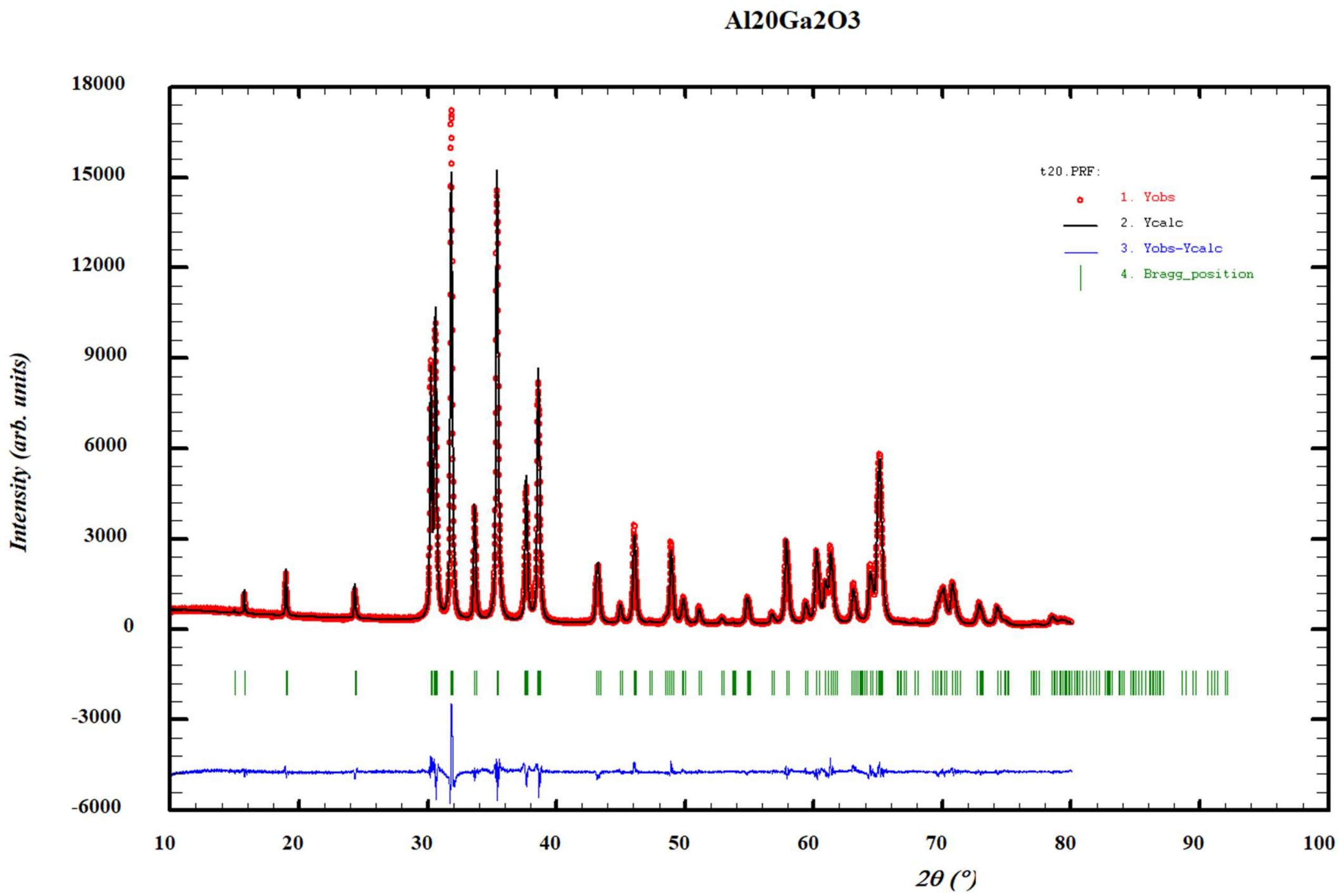

**Figure S4** Rietveld fitting of powder XRD data of β-($(Al_{0.2}Ga_{0.8})_2O_3$). File generated from WinPlotr. Symbols indicate experimental XRD data. Red line is the Rietveld fitted data. Green lines indicate the positions of Bragg peaks. Blue line is the difference between experimental and fitted data.

**Fig. S5**

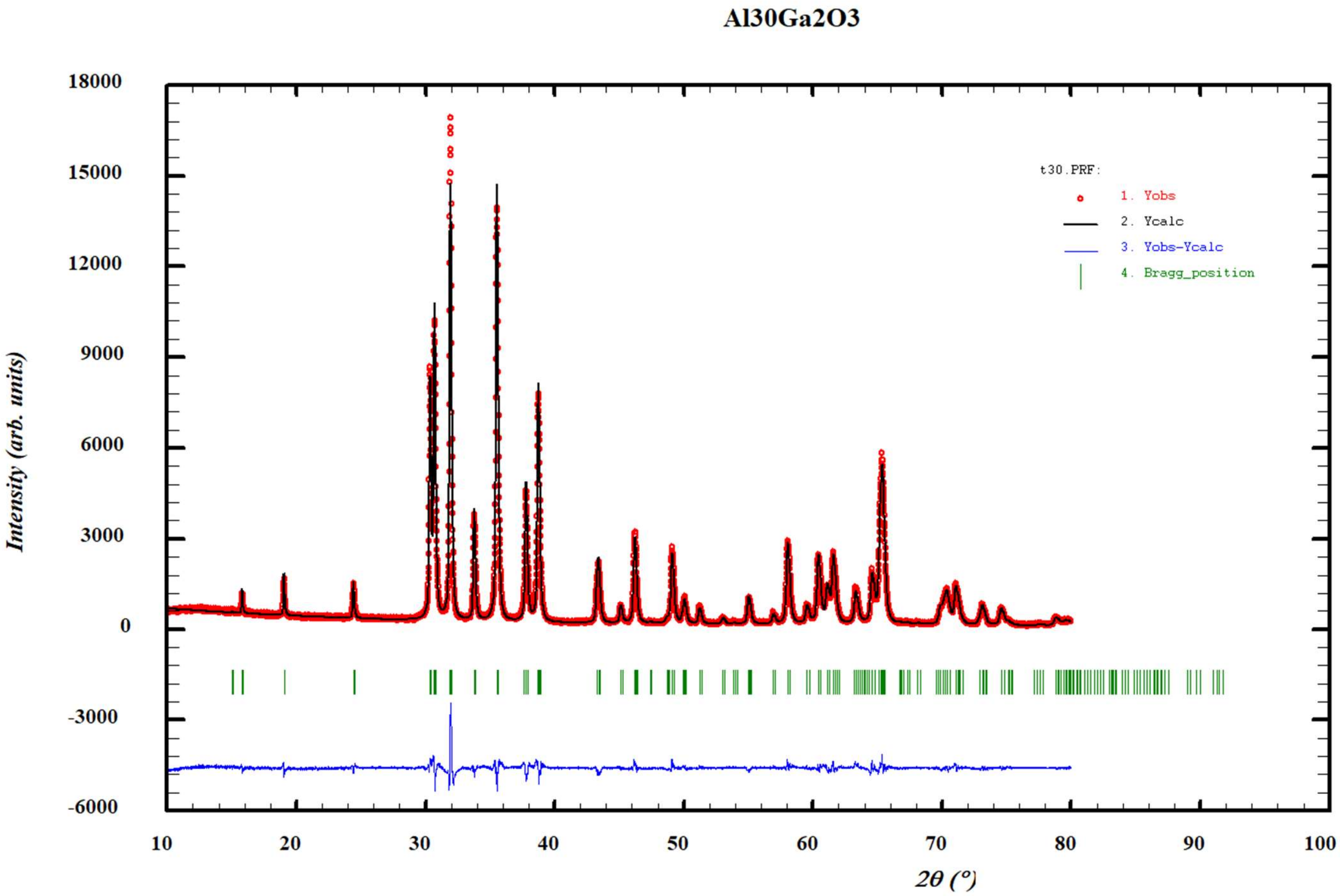

**Figure S5** Rietveld fitting of powder XRD data of β-$(Al_{0.3}Ga_{0.7})_2O_3$. File generated from WinPlotr. Symbols indicate experimental XRD data. Red line is the Rietveld fitted data. Green lines indicate the positions of Bragg peaks. Blue line is the difference between experimental and fitted data.

**Fig. S6**

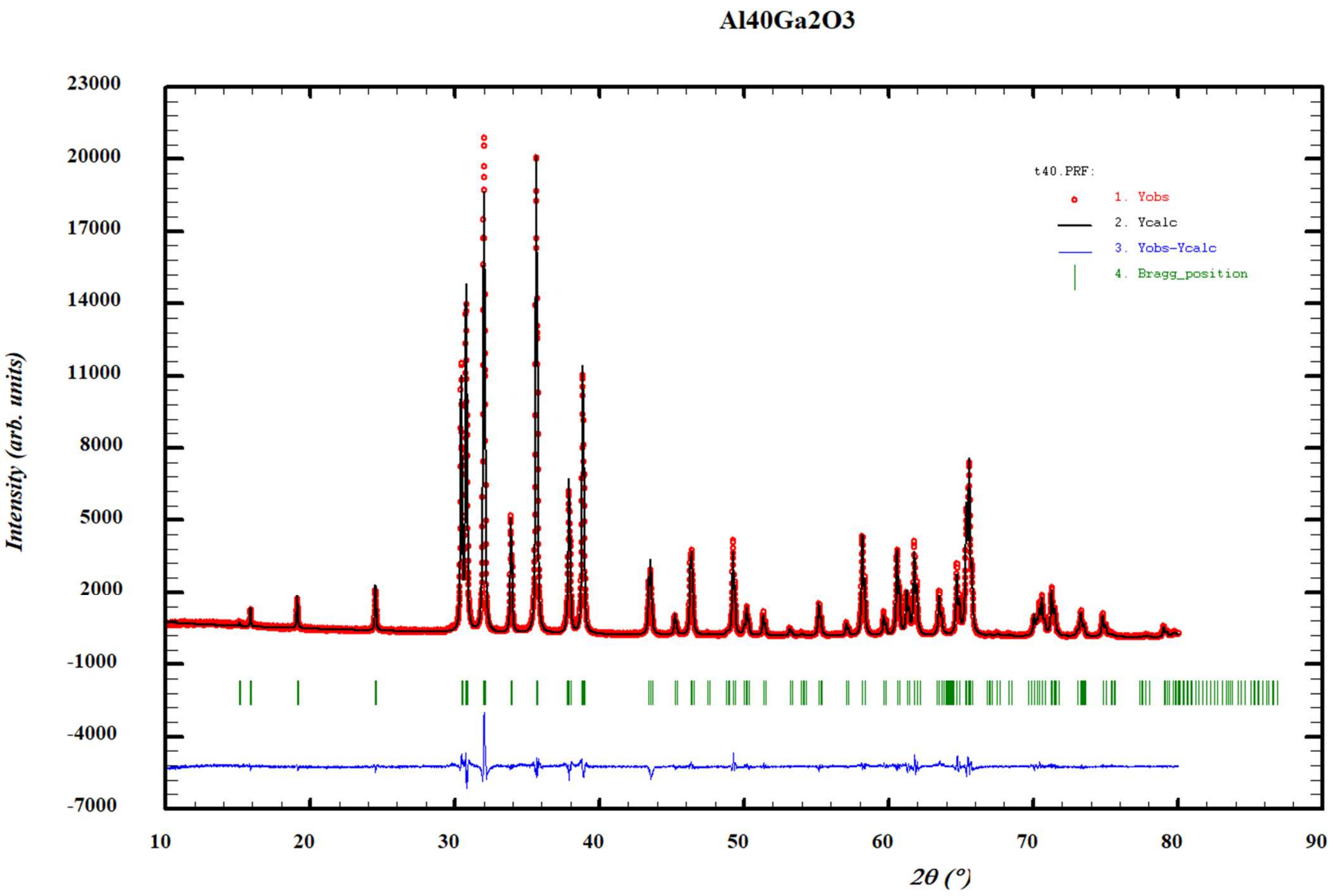

**Figure S6** Rietveld fitting of powder XRD data of β-$(Al_{0.4}Ga_{0.6})_2O_3$. File generated from WinPlotr. Symbols indicate experimental XRD data. Red line is the Rietveld fitted data. Green lines indicate the positions of Bragg peaks. Blue line is the difference between experimental and fitted data.

**Fig. S7**

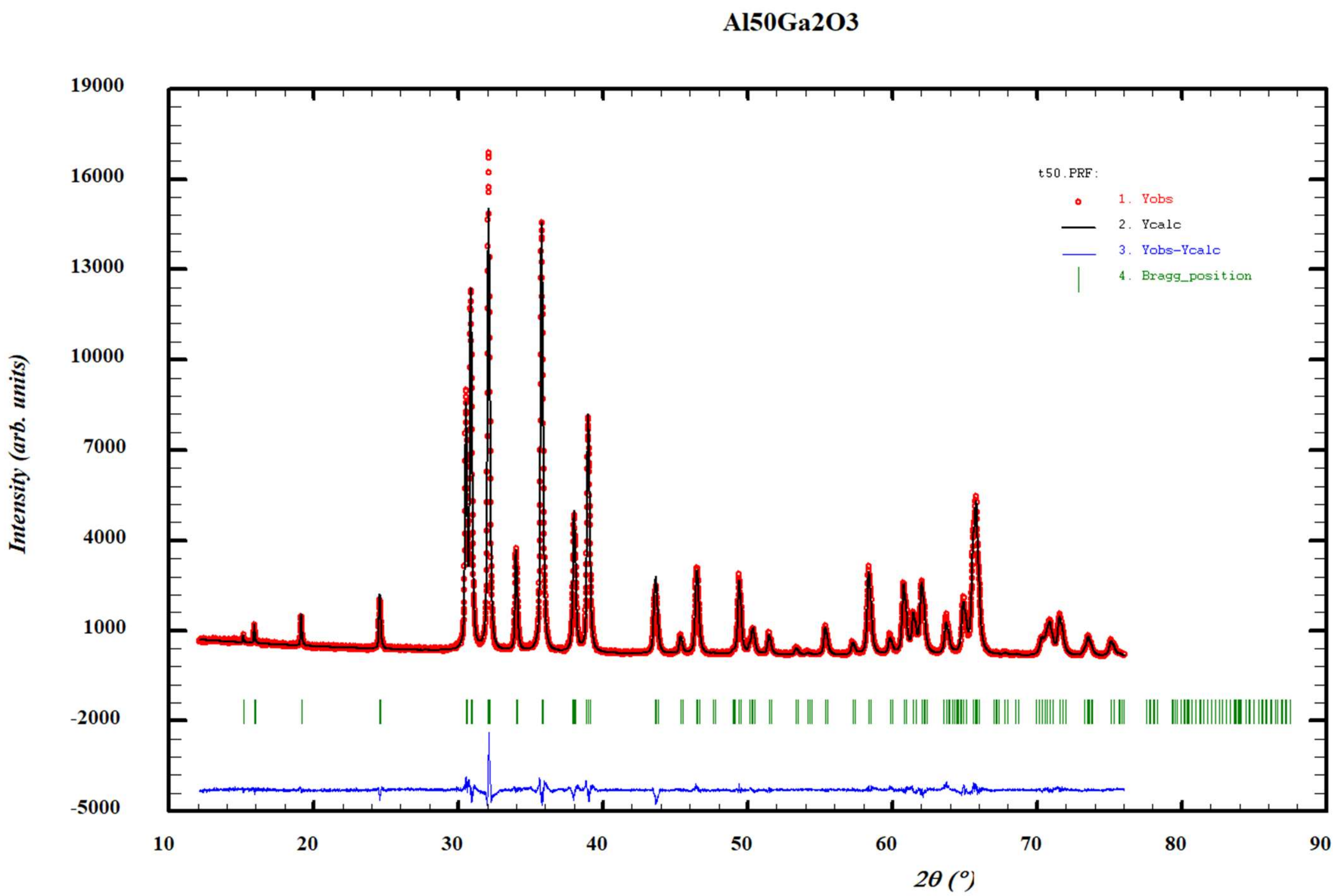

**Figure S7** Rietveld fitting of powder XRD data of β-$(Al_{0.5}Ga_{0.5})_2O_3$. File generated from WinPlotr. Symbols indicate experimental XRD data. Red line is the Rietveld fitted data. Green lines indicate the positions of Bragg peaks. Blue line is the difference between experimental and fitted data.

**Fig. S8**

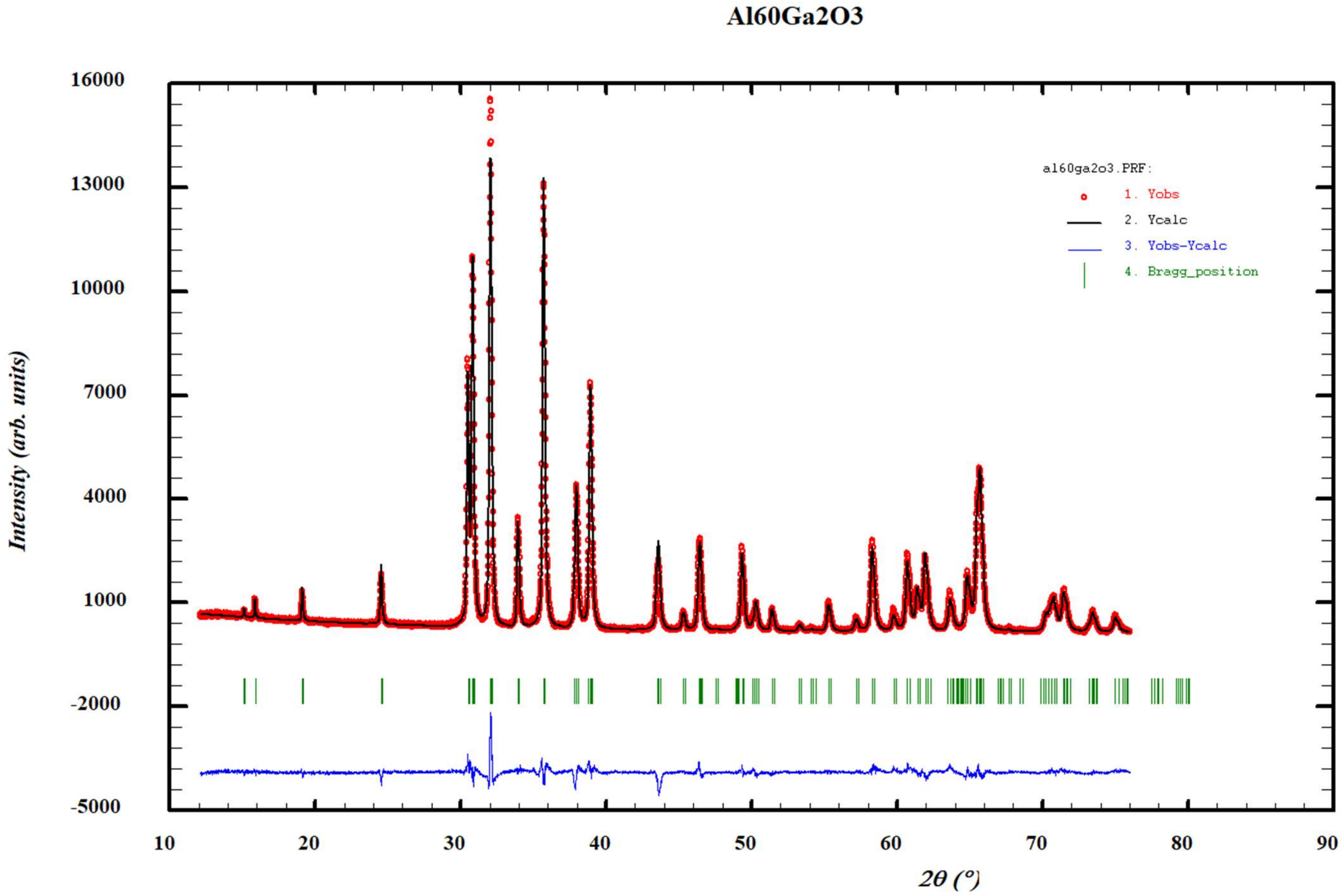

**Figure S8** Rietveld fitting of powder XRD data of β-$(Al_{0.6}Ga_{0.4})_2O_3$. File generated from WinPlotr. Symbols indicate experimental XRD data. Red line is the Rietveld fitted data. Green lines indicate the positions of Bragg peaks. Blue line is the difference between experimental and fitted data.

**Fig. S9**

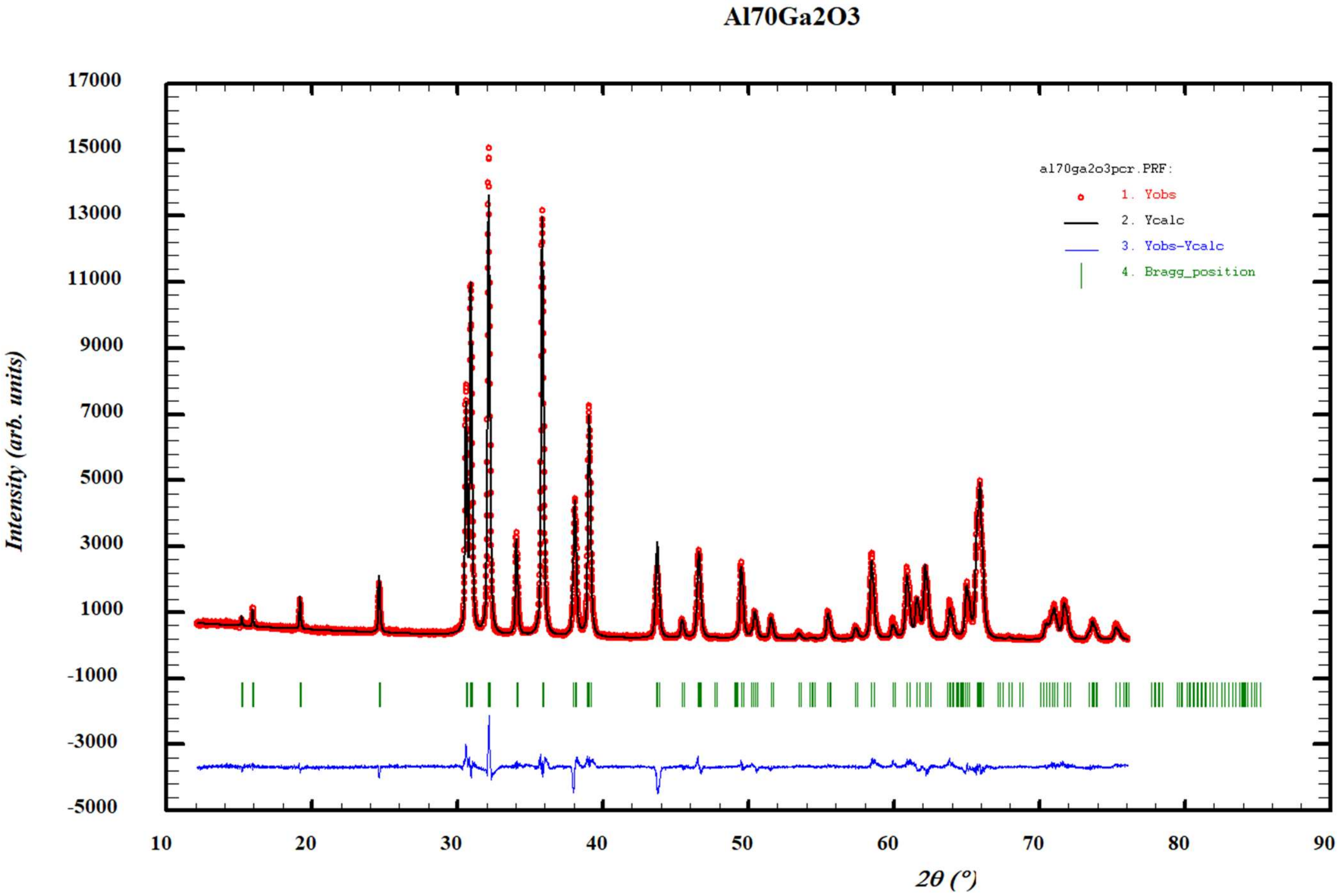

**Figure S9** Rietveld fitting of powder XRD data of β-$(Al_{0.7}Ga_{0.3})_2O_3$. File generated from WinPlotr. Symbols indicate experimental XRD data. Red line is the Rietveld fitted data. Green lines indicate the positions of Bragg peaks. Blue line is the difference between experimental and fitted data.

**Fig. S10**

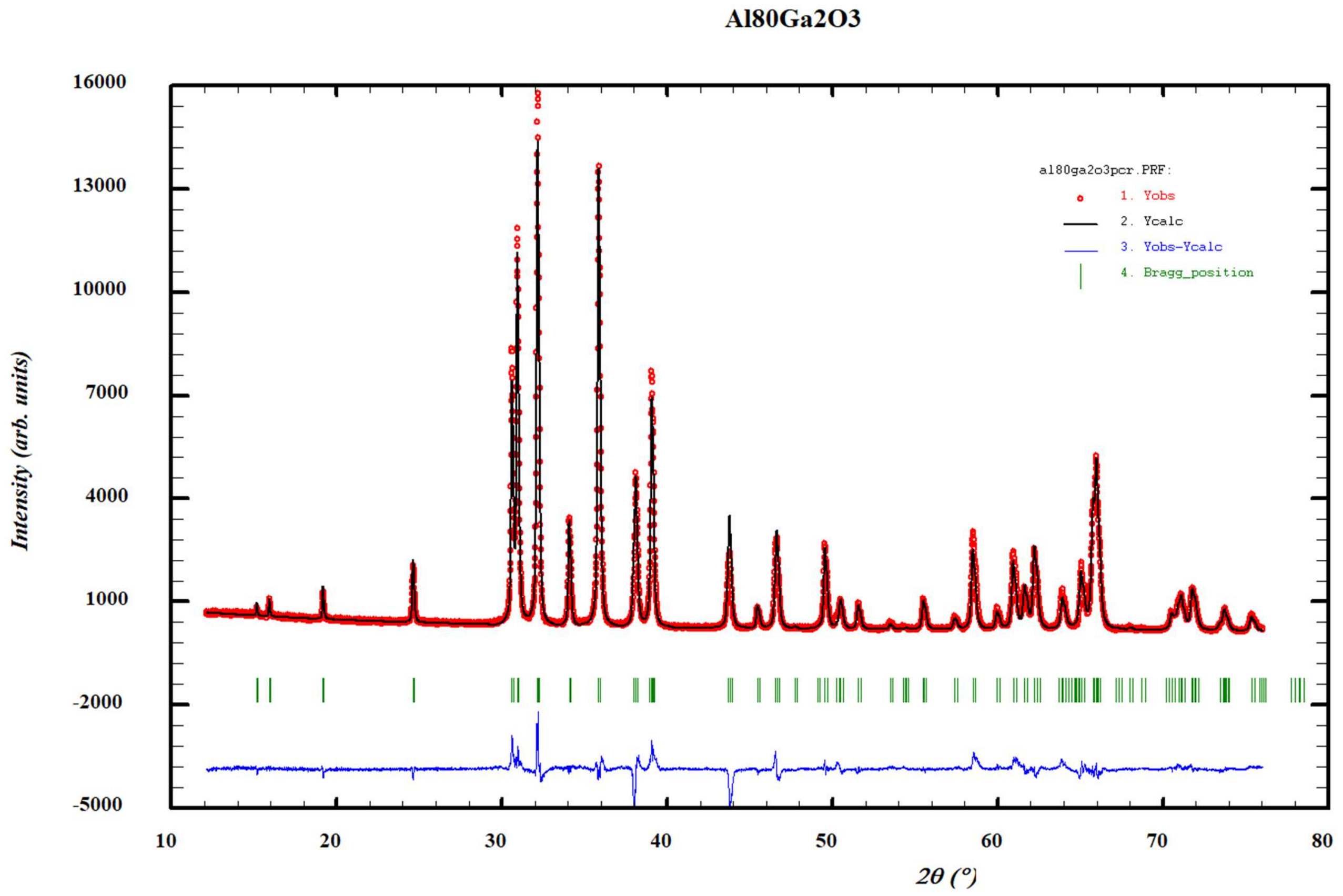

**Figure S10** Rietveld fitting of powder XRD data of β-$(Al_{0.8}Ga_{0.2})_2O_3$. File generated from WinPlotr. Symbols indicate experimental XRD data. Red line is the Rietveld fitted data. Green lines indicate the positions of Bragg peaks. Blue line is the difference between experimental and fitted data.

**Fig. S11**

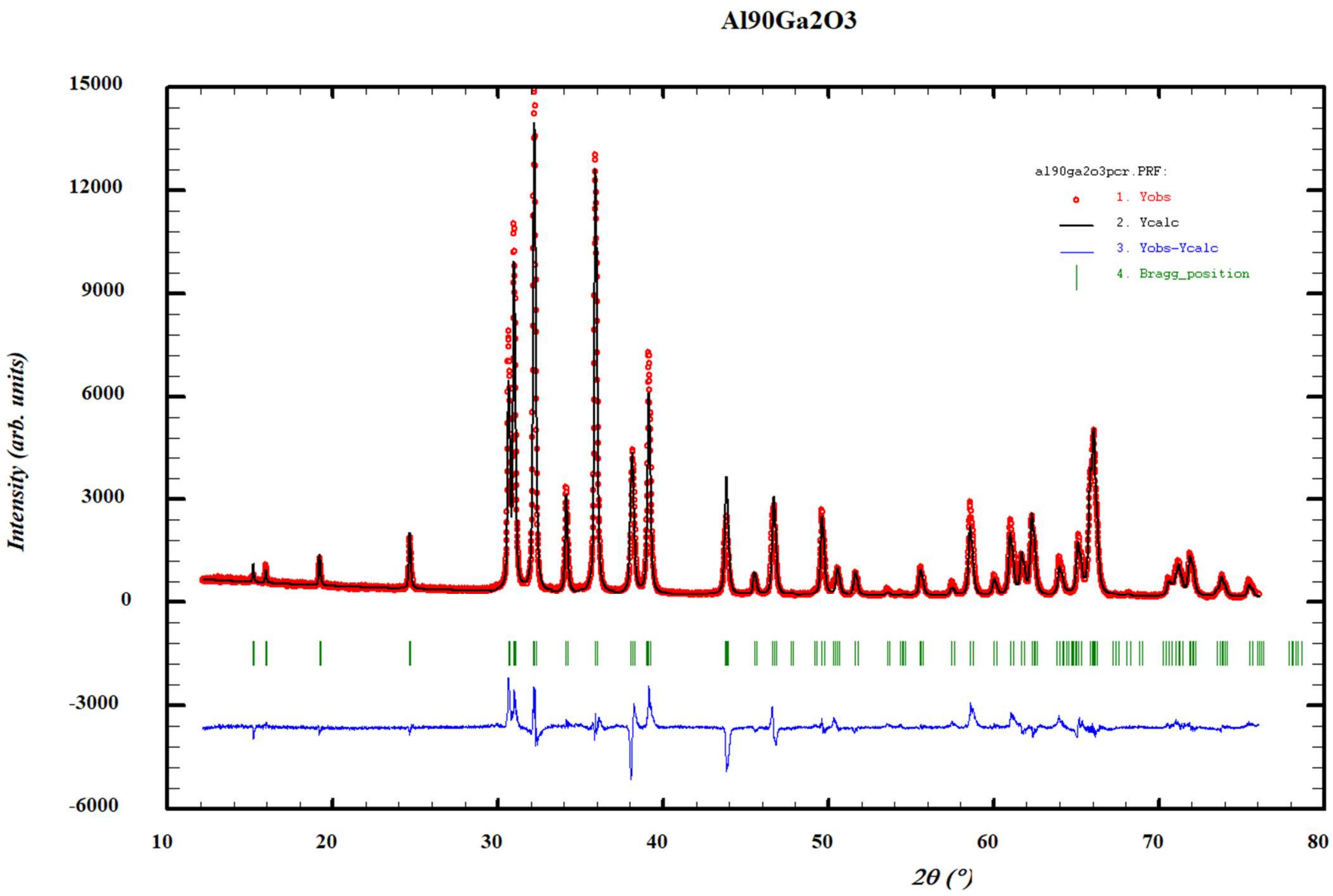

**Figure S11** Rietveld fitting of powder XRD data of β-$(Al_{0.9}Ga_{0.1})_2O_3$. File generated from WinPlotr. Symbols indicate experimental XRD data. Red line is the Rietveld fitted data. Green lines indicate the positions of Bragg peaks. Blue line is the difference between experimental and fitted data.